\documentclass{jpp}

\usepackage{graphicx}
\usepackage{natbib}
\usepackage{amsmath,amssymb}

\ifCUPmtlplainloaded \else
  \checkfont{eurm10}
  \iffontfound
    \IfFileExists{upmath.sty}
      {\typeout{^^JFound AMS Euler Roman fonts on the system,
                   using the 'upmath' package.^^J}%
       \usepackage{upmath}}
      {\typeout{^^JFound AMS Euler Roman fonts on the system, but you
                   dont seem to have the}%
       \typeout{'upmath' package installed. JPP.cls can take advantage
                 of these fonts, if you use 'upmath' package.^^J}%
      }
  \else
  \fi
\fi

\ifCUPmtlplainloaded \else
  \checkfont{msam10}
  \iffontfound
    \IfFileExists{amssymb.sty}
      {\typeout{^^JFound AMS Symbol fonts on the system, using the
                'amssymb' package.^^J}%
       \usepackage{amssymb}%

      }{}
  \fi
\fi

\ifCUPmtlplainloaded \else
  \IfFileExists{amsbsy.sty}
    {\typeout{^^JFound the 'amsbsy' package on the system, using it.^^J}%
     \usepackage{amsbsy}}
    {}
\fi

\newcommand{\be}{\begin{equation}}
\newcommand{\ee}{\end{equation}}
\newcommand{\bdm}{\begin{displaymath}}
\newcommand{\edm}{\end{displaymath}}
\newcommand{\bea}{\begin{eqnarray}}
\newcommand{\eea}{\end{eqnarray}}
\newcommand{\ba}{\begin{align}}
\newcommand{\ea}{\end{align}}

\newcommand{\varv}{v}

\newcommand{\ltsim}{\mbox{{\raisebox{-0.4ex}{$\stackrel{<}{{\scriptstyle\sim}}$}}}}

\newcommand{\araa}{ARA\&A}		
\newcommand{\apj}{ApJ}			
\newcommand{\apjl}{ApJLett}		
\newcommand{\aap}{A\&A}			
\newcommand{\grl}{Geophys.~Res.~Lett.} 
\newcommand{\jgr}{J.~Geophys.~Res.}	

\usepackage{graphicx}

\title
[Parametric decay of parallel and oblique Alfv\'en waves]
{Parametric decay of parallel and oblique Alfv\'en waves in the expanding solar wind}

\author[L.~{Del Zanna}, \emph{et al.}]
{L.~Del Zanna$^{1,2,3}$\thanks{luca.delzanna@unifi.it}, 
L.~Matteini$^{1,4}$, S.~Landi$^{1,2}$, A.~Verdini$^{1,5}$, M.~Velli$^{1,6}$}
\affiliation{
$^1$Dipartimento di Fisica e Astronomia, Universit\`a degli Studi di Firenze, Italy \\
$^2$INAF - Osservatorio Astrofisico di Arcetri, Firenze, Italy \\
$^3$INFN - Sezione di Firenze, Italy \\
$^4$Space and Atmospheric Physics Group, Imperial College London, UK \\
$^5$Solar--Terrestrial Center of Excellence, Royal Observatory of Belgium, Brussels, Belgium \\
$^6$Jet Propulsion Laboratory, California Institute of Technology, Pasadena, California, USA}

\pubyear{2014}
\volume{?}
\pagerange{1--18}
\date{?; revised ?; accepted ?. - To be entered by editorial office}

\begin{document}

\maketitle

\begin{abstract}
The long-term evolution of large-amplitude Alfv\'en waves propagating in the solar wind 
is investigated by performing two-dimensional MHD simulations
within the expanding box model.
The linear and nonlinear phases of the parametric decay instability are
studied for both circularly polarized waves in parallel propagation
and for arc-polarized waves in oblique propagation.
The non-monochromatic case is also considered. In the oblique case,
the direct excitation of daughter modes transverse to the local background field
is found for the first time in an expanding environment, and 
this transverse cascade seems to be favored for monochromatic mother waves.
The expansion effect reduces the instability growth rate, and it can even
suppress its onset for the lowest frequency modes considered here, possibly
explaining the persistence of these outgoing waves in the solar wind. 
\end{abstract}


\section{Introduction}

An exceptional laboratory for plasma physics and magnetohydrodynamics (MHD)
is represented by the solar wind, the supersonic continuous outflow
originating from the hot ($T>10^6 K$) corona of the Sun and permeating
the whole heliosphere.
The low-frequency part of fluctuations spectrum ($\nu\sim 10^{-4}-10^{-2}$~Hz) 
in high-speed and polar regions of the solar wind is known to be dominated by 
large-amplitude Alfv\'en waves propagating outwards \citep{Belcher:1971, Bruno:1985, 
Roberts:1987,  Grappin:1990, Marsch:1990, Tu:1990, Goldstein:1995, Horbury:2005}. 
The presence of such waves is also seen to affect the correlation between the wind speed
itself and the local field direction \citep{Matteini:2014}.
The ratio of incoming  to outgoing wave energies gradually increases and
saturates at heliocentric distances of about 2.5~AU  \citep{Bavassano:2000, Bruno:2013}, 
as derived from data from the \emph{Ulysses} spacecraft, which has probed the mostly uniform polar wind.
This behavior appears to contradict the so-called \emph{dynamical alignment} \citep{Dobrowolny:1980},
for which any initial imbalance in the Alfv\'enic modes should actually be reinforced by the
nonlinear interactions, at least within the framework of incompressible MHD.

However, in the solar wind plasma, compressibility effects are certainly likely to be influent.
An important consequence is that an Alfv\'en wave propagating on a background
magnetic field while preserving its (even large) amplitude, in spite of being
an exact solution of the full MHD equations, after some time becomes unstable because
of wave-wave coupling with compressible modes and decays into a backscattered
Alfv\'enic mode and a forward propagating acoustic mode \citep{Galeev:1963, Sagdeev:1969},
which are resonantly amplified. Thus, compressible modes are not only important for
their natural tendency to steepening, shock dissipation, and subsequent plasma heating, 
but also they may be responsible for the simultaneous presence of opposite propagating
Alfv\'enic modes, which is a necessary condition for the development of (incompressible) 
MHD turbulence.

Such \emph{parametric decay instability} is generally stronger in a low-beta plasma 
and for large amplitudes of the pump (or mother) Alfv\'en wave. 
Analytical results are concerned with the complete dispersion relation
 \citep{Derby:1978,Goldstein:1978} and some limiting regimes \citep{Hollweg:1994}.  
Numerical MHD simulations investigated
various aspects of this instability, from the dependence on the plasma beta,
to the propagation in multi-dimensions, to the different polarizations, to the
presence of  an initial broad-band spectrum of fluctuations \citep{Hoshino:1989,Vinas:1991,
Umeki:1992, Ghosh:1993, Malara:1996, Malara:2000, Del-Zanna:2001, Del-Zanna:2001a}.
In these latter papers it was shown that parametric decay can indeed be responsible
for the observed gradual increase of incoming Alfv\'enic flux with distance, or correspondingly
the reduction of the normalized \emph{cross helicity} $\sigma_c$
\citep[see][for a critical discussion]{Inhester:1990}, 
whereas its saturation at a certain distance can be easily explained by 
the natural suppression of the decay instability itself when the nonlinear regime is reached.

In particular, \citet{Del-Zanna:2001} showed that the saturation of the instability
always occurs when the daughter acoustic mode has steepened into a train of quasi-parallel shocks,
that may eventually contribute to the plasma heating. This may be important
in solar coronal holes, where the plasma beta is lower and the decay more
efficient (decay time and length-scales are shorter and up to half of the initial Alfv\'enic 
energy is observed to go into heat), and where the lack of large-scale magnetic structures 
suggests to exclude other mechanisms \citep{Del-Zanna:2002a}.
However, even at larger heliocentric distances, where the beta is higher
and also the fluid description should break down in favor of a kinetic regime,
hybrid simulations of parametric instabilities show plasma heating via particle 
trapping in the potential wells associated with the steepened resonant density perturbation 
\citep{Nariyuki:2006,Nariyuki:2007,Araneda:2008,Matteini:2010a,Matteini:2010}. 
Moreover, proton velocity beams are seen to develop in the distribution functions,
reminiscent of what is observed in real data.

Another important issue is that the majority of the observed large-amplitude Alfv\'en waves 
in the solar wind are found with a spherical \emph{arc-like} polarization \citep{Riley:1996,Tsurutani:1999}.
This means that one should not restrict the analysis to circularly polarized waves
in parallel propagation, with respect to the average background magnetic field, 
but it is important to study the more general case of oblique propagation. 
Numerical investigations \citep{Vasquez:1996,Del-Zanna:2001a}
have shown that the basic properties of the instability are preserved, and that in addition
this can also lead to a \emph{direct} creation of transverse small-scales 
\citep{Matteini:2010a, Del-Zanna:2012}.
This possibility may provide a complementary mechanism to (or at least a seed for) 
strong Alfv\'enic turbulence \citep{Breech:2005}. Notice that the simultaneous
presence of compressible heating and transverse Alfv\'enic turbulence
appears to be required in stationary models for fast solar wind acceleration 
\citep{Verdini:2007,Verdini:2010}. See however \citet{Lionello:2014} for
a comparison with a time-dependent model.

In order to properly study the nonlinear wave or turbulence
evolution in the solar wind, the effects of the radial expansion
should be taken into account. 
Cartesian-like periodic domains are usually employed for such kind
of simulations, here we adopt the so-called \emph{expanding box} (EB) model
\citep{Velli:1992, Grappin:1993, Grappin:1996}, also adapted to kinetic and hybrid models 
\citep{Liewer:2001, Hellinger:2005, Matteini:2006}.
The radial expansion effects are mimicked by introducing a lateral
stretching of a local frame comoving with the solar wind velocity.
In this model, effects like the gradual decrease in wave amplitude
and the development of the Parker spiral are easily recovered.

The EB model has been recently implemented in the 
\emph{Eulerian Conservative High Order} ({\tt ECHO}) code
for classical and relativistic MHD \citep{Del-Zanna:2007, Landi:2008, Del-Zanna:2009a},
taking advantage of the freedom in the choice of the spatial metric.
Preliminary test simulations \citep{Del-Zanna:2012a} have shown that  
the expansion is likely to affect the instability growth and even to prevent
it, depending on the respective characteristic time-scales.
A one-dimensional investigation of the effects of a non-uniform wind velocity
on parametric decay has also been recently performed by \citet{Tenerani:2013}.
This should be appropriate for coronal holes and for the wind accelerating region
where the flow may still be sub-Alfv\'enic.

In the present paper we perform two-dimensional MHD simulations
by applying the EB model to circularly 
polarized Alfv\'en waves in parallel propagation, and to
arc-polarized Alfv\'en waves in oblique propagation, to study the
direct excitation of perpendicular modes. Both monochromatic
mother waves or an initial broad-band spectrum will be considered.
The effect of the solar wind expansion will be investigated by
changing the location of the initial position of the comoving box.
The paper is organized as follows. The EB-MHD equations and their implementation
in the {\tt ECHO} code are described in Sect.~\ref{sect:model}, the initial conditions and
setup relevant for the simulations of parametric decay are in Sect.~\ref{sect:init},
simulation results will be shown in Sect.~\ref{sect:results}, while conclusions will be given in
Sect.~\ref{sect:conclusions}.

\section{The Expanding Box model for MHD}
\label{sect:model}

When studying numerically wave motion or any kind of plasma instabilities,
it is more convenient, whenever possible, to use a local simulation box in
Cartesian-like coordinates and periodic boundary conditions. Within such settings
the physical effect under investigation can be more easily singled out
and high-accuracy numerical methods employed. In the present section
we first describe the original version of the EB model, leading to a few additional source
terms in the set of MHD equations, and a novel one, more appropriate for
multi-purpose conservative schemes working in any system of coordinates.

In the case of the solar wind, it is useful to
consider a parcel of plasma in the radially expanding outflow, characterized
by a given background velocity profile $U(R)$.  If we identify
the heliocentric distance of the parcel in time with
$R(t)$, so that $dR/dt\equiv U(R)$, it is convenient to introduce, 
in analogy with cosmology, the \emph{expansion scale factor} 
\be
a(t) = \frac{R(t)}{R_0}, \quad \frac{\dot{a}}{a} = \frac{U(R)}{R},
\label{eq:a}
\ee
where $R_0\equiv R(0)$ is the initial parcel position, the dot indicates time derivation, 
and the latter quantity in Eq.~(\ref{eq:a}) is the inverse of the expansion characteristic time,
basically the analog of the cosmological Hubble time.

Let us now choose a locally Cartesian-like coordinate system
with $x=r-R(t)$ and \emph{stretched} transverse coordinates $y$ and $z$,
expanding precisely at the rate $a(t)$ due to the radially diverging geometry.
Provided the box is taken sufficiently small such that the wind speed can
be considered as uniform, the latter can be removed with a Galilean transformation.
In the new \emph{comoving} frame $x^1=x,\,x^2=y/a,\,x^3=z/a$
the residual velocity is, to first-order expansion
\be
\mathbf{u}_\perp = \dot{a}\,(0, x^2, x^3)=(\dot{a}/a) (0, y, z),
\ee
clearly non-vanishing in the transverse direction. 
In the evolution equations, time derivatives and (transverse) spatial gradients
are therefore modified as
\be
\partial_t \ \rightarrow \partial_t  - (\mathbf{u}_\perp \cdot \mathbf{\nabla}), \quad
\mathbf{\nabla} = (\partial_1, \, a^{-1}\partial_2, \, a^{-1}\partial_3)=(\partial_x,\partial_y,\partial_z),
\ee
so that novel source terms are expected to appear.

The system of ideal MHD equations in the EB approximation is \citep{Grappin:1993}:
\be
\partial_t \rho + \mathbf{\nabla}\cdot (\rho \mathbf{\varv} ) =  - (\dot{a}/a) 2\rho ,
\ee
\be
\rho ( \partial_t + \mathbf{\varv}\cdot\mathbf{\nabla} ) \mathbf{\varv}  - (\mathbf{B}\cdot\mathbf{\nabla}) \mathbf{B} + \mathbf{\nabla} (p + B^2/2)
 =  - (\dot{a}/a) \rho \mathbf{v}\cdot \mathcal{T} ,
\ee
\be
( \partial_t + \mathbf{\varv}\cdot\mathbf{\nabla} ) p + \Gamma p \mathbf{\nabla} \cdot\mathbf{\varv} =
 - (\dot{a}/a) 2\Gamma p,
\ee
\be
 \partial_t \mathbf{B} - \mathbf{\nabla}\times (\mathbf{\varv}\times\mathbf{B})  =
  - (\dot{a}/a) \mathbf{B}\cdot (2\mathcal{I} - \mathcal{T}).
\ee
In the above equations $\rho$ is the density, $\mathbf{\varv}$ is the velocity in the comoving frame, 
$p$ is the thermal pressure, $\mathbf{B}$ is the magnetic field (we use Gaussian 
units with $4\pi\to 1$), $\mathcal{I} $ is the identity tensor, $\mathcal{T}=\mathrm{diag}\{0,1,1\}$, 
and $\Gamma = 5/3$ is the adiabatic index for an ideal monoatomic gas.

The above ideal EB-MHD equations predict, even for a static and uniform plasma, an evolution
$\rho\sim a^{-2}$, $T\sim p/\rho\sim a^{-2(\Gamma-1)}\sim a^{-4/3}$, $B_x\sim a^{-2}$, $B_y\sim a^{-1}$, 
$B_z\sim a^{-1}$, just due to the lateral stretching (notice the formation of the Parker spiral,
in the $x-y$ plane, as $\tan\theta = B_y/B_x\sim a$). 
If Alfv\'enic fluctuations are present, in the strong coupling regime the WKB approximation
for small amplitudes and short wavelengths predicts that $\delta\varv^2\sim\delta B^2/\rho\sim a^{-1}$.
The model of the expanding box  is thus able to capture all known scaling laws
with heliocentric distance.

In order to study numerically the nonlinear evolution of Alfv\'en waves, it is convenient
to adopt the conservative formulation for the MHD equations. In particular, the {\tt ECHO} code
solves for the classical (and relativistic) MHD equations in the general form
\be
\partial_t \mathcal{U} + \partial_i \mathcal{F}^i = \mathcal{S},
\ee
where $\mathcal{U}$ is a generic vector of conservative variables,
$\mathcal{F}^i$ are the corresponding fluxes (latin indexes like $i$ run
on the 3 spatial coordinates $x^i$ and the Einstein convention of a sum over
repeated indexes is implicitly assumed), 
and $\mathcal{S}$ is the vector containing the source terms.
Given a generic spatial metric tensor $g_{ij}$, with determinant $g$,
the MHD system is retrieved by choosing 
\be
\mathcal{U} \! = \!   \sqrt{g}\left[\begin{array}{c}
\rho \\ \rho \varv_j \\ E_t \\ B^j \end{array}\right], \quad
\mathcal{F}^i \! = \!  \sqrt{g} \left[\begin{array}{c}
\rho \varv^i \\ \rho \varv^i\varv_j - B^iB_j + p_t \delta^i_j \\
(E_t + p_t) \varv^i - (\varv_kB^k)B^i \\ \varv^i B^j - \varv^j B^i \\
\end{array}\right],  \,
\ee
where the \emph{total} kinetic plus magnetic pressure and energy are
\be
p_\mathrm{t}=p+\textstyle{\frac{1}{2}}B^2,\quad
E_\mathrm{t}= \textstyle{\frac{1}{2}}\rho \varv^2 + 
\textstyle{\frac{1}{\Gamma -1}}\,p + 
\textstyle{\frac{1}{2}}B^2,
\ee
and the source terms are
\be
\mathcal{S}= \sqrt{g}  \left[\begin{array}{c}
0 \\  \textstyle{\frac{1}{2}} (\rho \varv^i\varv^k - B^iB^k + p_t g^{ik})\partial_j g_{ik} \\  
- \frac{1}{2}(\rho \varv^i\varv^j - B^iB^j + p_t g^{ij})\partial_t g_{ij}\\ 0
\end{array}\right].
\ee
Obviously, for non-Cartesian metrics, care must be taken 
to the differences between covariant and contravariant components 
(related by $\varv_i=g_{ij}\varv^j$).

It is now straightforward to introduce EB within such framework. 
Given the scale factor $a=a(t)$ of Eq.~\ref{eq:a} yielding the transverse expansion, 
the metric tensor must be necessarily defined as
\be
g_{ij}=\mathrm{diag}\{1,a^2(t),a^2(t)\},\quad \sqrt{g}=a^2(t),
\label{metric}
\ee
thus the metric is almost Cartesian (diagonal and spatially uniform),
but time-dependent. If we want to employ the usual Cartesian orthonormal 
$x$, $y$, and $z$ components, we must recall that 
for a generic vector $\varv^i$ we have $\varv_1=\varv^1=\varv_x$, 
$\varv_2=a^2\varv^2=a\varv_y$, $\varv_3=a^2\varv^3=a\varv_z$.
Notice that the source term in the momentum equation vanishes 
($\partial_j g_{ik}=0$), while that in the energy equation is given by
\be
- \tfrac{1}{2}(\rho \varv^i\varv^j - B^iB^j + p_t g^{ij})\partial_t g_{ij} = 
- \frac{\dot{a}}{a}(\rho \varv_\perp^2 - B_\perp^2 + 2p_t),
\ee
where $\varv^2_\perp=\varv_y^2+\varv_z^2$ and similarly for the magnetic field.
The original non-conservative EB-MHD equations described previously
are equivalent to the form above required by the {\tt ECHO} code, as it can be
easily verified (recall that $\partial_1=\partial_x,\,\partial_2=a\partial_y,\,\partial_3=a\partial_z$). 

In the present paper we will only consider the evolution of Alfv\'en waves
at large heliocentric distances, where the outflow is super-Alfv\'enic and
$U(R)$ can be considered as constant.  For an investigation appropriate to the
solar wind acceleration region see \citet{Tenerani:2013},
where the relevant modifications to the equations can also be found.
Here we thus choose for simplicity $U(R)\equiv U_0$ and
\be
R(t) = R_0 + U_0 t, \quad a(t) = 1 + \epsilon t, \quad 
\frac{\dot{a}}{a} = \frac{\epsilon}{1 + \epsilon t},
\ee
where $\epsilon \equiv U_0/R_0$ is the \emph{constant} expansion rate,
or, equivalently, the inverse of the characteristic expansion time.

\section{Simulation setup}
\label{sect:init}

The quantities appearing in the MHD equations previously described,
and in the initial conditions of the present section, are normalized
against values at the reference distance $R_0\equiv R(0)$ at simulation time $t=0$.
We choose Alfv\'enic units throughout the paper: a density $\rho_0$, a background magnetic field
with strength $B_0$, a basic wave period $\tau_\mathrm{A}$, so that the Alfv\'en speed 
is $\varv_\mathrm{A} = B_0/\sqrt{\rho_0}$, and any length will be expressed in terms 
of the basic wavelength $\lambda_\mathrm{A} = \varv_\mathrm{A}\tau_\mathrm{A}$.
At the initial position $R_0$ and time $t=0$ we also define the
plasma beta $\beta = c^2_\mathrm{s}/\varv^2_\mathrm{A}$, where
$c^2_\mathrm{s} = \Gamma p_0/\rho_0$ is the square of the sound speed.

The initial conditions are provided as follows. Once the values of the
plasma beta  $\beta$ and of the normalized mother wave amplitude
$\eta$  have been chosen, we set up an unperturbed 
static plasma with density $\rho=1$, pressure $p = \beta/\Gamma$, 
velocity $\mathbf{\varv}=0$, background field
\be
\mathbf{B}_0  = \cos\theta_0 \mathbf{e}_x + \sin\theta_0 \mathbf{e}_y,
\ee
where $\theta_0$ is the initial angle of the Parker spiral at $R_0$, and
Alfv\'enic fluctuations
\be
\delta\mathbf{B}  = \eta \cos (\varphi) \mathbf{e}_z + 
\mathcal{F}(\varphi) \mathbf{e}_y,
\quad \delta{\mathbf{\varv}} = -  \delta\mathbf{B},
\ee
where $\varphi = \mathbf{k}_0\cdot\mathbf{x}\equiv k_0 x$ is the phase for the
\emph{monochromatic} case (see however Sect.~\ref{sect:nonmon}),
$\mathbf{k}_0 = k_0 \mathbf{e}_x$, with $k_0=2\pi$, is the wave vector 
of the pump (mother) wave.

The function $\mathcal{F}$ is derived by the condition 
$|\mathbf{B}_0 + \delta\mathbf{B}|=\mathrm{const}$ and depends on the 
chosen polarization, as described below.
\begin{itemize}
\item \emph{Circular polarization (parallel propagation)} -- the oscillations are
always orthogonal to both $\mathbf{k}_0$ and $\mathbf{B}_0$, taken to be radial 
with $\theta_0=0$, and
\be
\mathcal{F}(\varphi) = \eta \sin (\varphi);
\ee
\item \emph{Arc polarization (oblique propagation)} -- in this case
$\mathbf{k}_0$ and $\mathbf{B}_0$ are not parallel and $B_y$ has both
a background and an oscillating part. By imposing $C^2=B_y^2 + B_z^2= B^2-{B_0}_x^2$
and $<\delta B_y >=0$ \citep{Barnes:1974,Del-Zanna:2001a} we find
\be
\mathcal{F}(\varphi) = \sqrt{C^2 - \eta^2 \cos^2 (\varphi)} - \sin\theta_0,
\ee
where the unknown constant $C$ is derived by solving the elliptic integral
\be
\frac{2}{\pi}\int_0^{\pi/2} \!\! \sqrt{C^2 - \eta^2 \cos^2 \varphi} \, \mathrm{d}\varphi = \sin\theta_0,
\ee
which admits solutions only if $\sin\theta_0 > 2\eta/\pi$.
\end{itemize}

The pump Alfv\'en waves in either circular or arc polarization described above
are \emph{exact} solutions of the ideal MHD equations (without expansion),
and propagate with 
\be
\omega_0 = \mathbf{k}_0\cdot\mathbf{\varv}_\mathrm{A}=k_0\cos\theta_0,
\ee
preserving $B=|\mathbf{B}_0+\delta\mathbf{B}|$ and $|\delta\mathbf{\varv}|$, 
even if the amplitude $\eta$ is large, constant in time during the evolution.
A last possibility is of course to abandon the condition of a constant $B$
for a simpler \emph{linear} polarization with $\mathcal{F}(\varphi)=0$,
either in parallel ($\theta_0=0$) or oblique propagation, but the evolution
is affected by ponderomotive pressure forces proportional to $\eta^2$,
and will not be considered here.

Parametric decay occurs when compressibility is taken into account.
Small-scale density perturbations may grow resonantly due to
three-modes wave-wave nonlinear interactions and the energy
of the pump wave is partially transferred to two daughter waves.
For quasi-normal modes (small amplitudes), and in the low-beta regime,
the mother wave ($\omega_0=k_0$) decays into a forward propagating 
sound wave ($\omega_c=\sqrt{\beta}k_c$) and a back-scattered Alfv\'en wave 
($\omega^-=k^-$), where parallel propagation has been considered for simplicity.
Under these approximations, the resonance condition is found by the
scattering relations $\omega_0=\omega_c + \omega^-$ and $k_0=k_c-k^-$,
leading to $k_c/k_0=2/(1+\sqrt{\beta})$ and $k^-/k_0=(1-\sqrt{\beta})/(1+\sqrt{\beta})$.
The growth rate of the instability can be approximated as \citep{Galeev:1963}
$\gamma/\omega_0 \simeq \eta\beta^{-1/4}/2$, so it is stronger for a low-beta
plasma and for large amplitudes.
For generic values of $\eta$ and $\beta$, that is even for large amplitudes and
warm plasmas, one needs to solve the full
dispersion relation. For circularly polarized waves and parallel propagation
it reads \citep{Derby:1978,Goldstein:1978}:
\be
(\omega - k)(\omega^2-\beta k^2)[(\omega+k)^2-4]=
\eta^2 k^2 (\omega^3 +k\omega^2 -3\omega +k),
\label{eq:dispersion}
\ee
where $k/k_0\to k$ and $\omega/\omega_0\to \omega$, looking for the range
of unstable modes. For instance, if we consider $\eta=0.2$ and $\beta=0.1$, the resulting
maximum growth rate is $\gamma/\omega_0\simeq 0.11$, obtained for
the compressive mode with wave number $k_c/k_0\simeq 1.55$.
In the oblique case, if retaining the 1D approximation $\mathbf{k}_0 = k_0 \mathbf{e}_x$,
these numbers scale as $\cos\theta_0$, since $\omega_0=k_0\cos\theta_0$,
though the parallel case can not be retrieved for $\theta_0=0$  \citep{Del-Zanna:2001a}.


\begin{figure}
\centerline{
 \includegraphics[height=5.0cm]{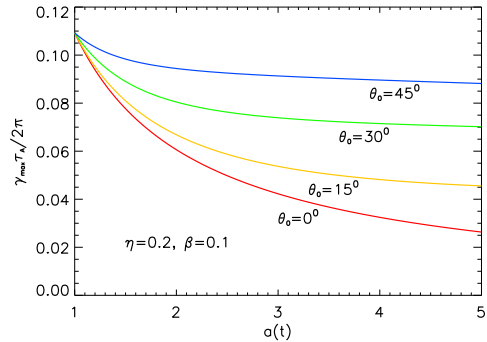}
  }
\caption{
The maximum growth rate for the decay instability as a function of the
expansion scale factor $a=R(t)/R_0$, for $\eta=0.2$, $\beta=0.1$,
and circular polarization.
Different values of the initial angle $\theta_0$ in the equatorial plane are compared.
}
\label{fig:dispersion}
\end{figure}


However, during the evolution in the expanding solar wind plasma the parameters
$\omega_0$, $\eta$, and $\beta$ are all dependent on $R$ and thus also
time-dependent with $a(t)$. For parallel propagation we have $\omega_0\sim a^{-1}$,
$\eta\sim a^{1/2}$, and $\beta\sim a^{2/3}$, so that the growth rate is expected
to decrease in time as $\gamma\sim a^{-2/3}$ \citep{Tenerani:2013}.
On the other hand, if the development of the Parker spiral is taken into account, that
is when $\theta_0>0$, the magnitude of the background field should
decrease initially as $B\sim a^{-2}$ (when $B_x$ is dominant) and after
a certain radius as $B\sim a^{-1}$ (when $B_y$ is dominant), so that
$\omega_0$ becomes constant while $\eta\sim a^{-1/2}$ and $\beta\sim a^{-1/3}$ 
start to decrease with time (and distance). The growth rate of the parametric
instability is now expected to decrease as $\gamma\sim a^{-5/12}$.
The above estimates are valid when the expansion rate is small 
compared to the growth rate, and for small values of $\eta$ and $\beta$.
For the most general case of warm plasmas and large amplitudes, still preserving
the condition $\epsilon\ll\gamma$, one has instead to solve 
Eq.~\ref{eq:dispersion} with input values rescaled by the expansion factor. 
Results are plotted in Fig.~\ref{fig:dispersion}, where the maximum rate
as a function of $a(t)$ is plotted for different values of $\theta_0$.

Simulations will be performed in the ecliptic plane $x-y$ by using 
a two-dimensional numerical domain $[0, L_x]\times [0,L_y]$, 
where $L_x = L_y = m_0\lambda_\mathrm{A}$ at the
initial time, and $L_y\sim a$ will be stretched due to solar wind radial expansion,
as a built-in effect characteristic of EB. Notice that,
due to expansion, not only the maximum growth rates diminish but also
the correspondent wave numbers. Thus, $m_0$
must be sufficiently high in order to have room for daughter waves to fully develop
in our numerical domain with periodical boundary conditions, especially in the oblique case
where transverse modes are expected to be excited too.  We choose throughout
the paper the value $m_0=10$.

Finally, we must choose reference values for our parameters and
ensure that the EB approximation is applicable. The normalized
expansion parameter is
\be
\epsilon =
5.0 \times 10^{-2} \left(\frac{U_0}{750~\mathrm{km~s}^{-1}}\right) 
\left(\frac{R_0}{1~\mathrm{AU}}\right)^{-1}
 \left(\frac{\tau_\mathrm{A}}{10^4~\mathrm{s}}\right),
 \label{eq:epsilon}
\ee
increasing for smaller values of the initial radial distance of the box $R_0$.
We choose $R_0=0.25~\mathrm{AU}$, so that the evolution will be limited in the range 
where \emph{in situ} data are available, from the Helios mission in the inner region up to
the Earth's orbit and beyond. For fast streams in the ecliptic plane at that distance
we assume the wind velocity to be already in its asymptotic regime with 
$U_0=750~\mathrm{km~s}^{-1}$. If $\tau_\mathrm{A}=10^4~\mathrm{s}$ 
we find the maximum expansion rate of $\epsilon=0.2$. Moreover,
assuming a dependence of $\propto 1/r$ for the Alfv\'en speed (we use
the value $\varv_\mathrm{A}=50~\mathrm{km~s}^{-1}$ at 1~AU), we are able
to estimate the maximum length for the numerical box, which must satisfy the rule
\be
\frac{L_x/m_0}{R_0} \ll 3.3\times 10^{-3}
 \left(\frac{\varv_\mathrm{A}}{50~\mathrm{km~s}^{-1}}\right) 
 \left(\frac{R_0}{1~\mathrm{AU}}\right)^{-2}
 \left(\frac{\tau_\mathrm{A}}{10^4~\mathrm{s}}\right),
\ee
that is the Alfv\'en wavelength $\lambda_\mathrm{A} = \varv_\mathrm{A}\tau_\mathrm{A}$
($m_0$ modes in the box of length $L_x$) must be
smaller than the scale of the background gradients, of the order of $R_0$.

When $R_0=0.25~\mathrm{AU}$ and $m_0=10$ we see that $L_x \approx 0.5 R_0$,
so that the maximum period that we are able to study is indeed $\tau_\mathrm{A}=10^4~\mathrm{s}$,
whereas higher frequencies would be much better suited.
In our simulations we will investigate the Alfv\'enic period $\nu_\mathrm{A}=10^{-4}-10^{-2}$~Hz,
as observed at Helios distances, corresponding to an \emph{injection} spectrum $\nu^{-1}$ of modes, 
presumably still of solar origin, that we will use as pump waves for the decay instability.

\section{Results}
\label{sect:results}

\begin{table}
\caption{List of simulation parameters}
\centering
\begin{tabular}{l l l l l l l l l l}
\hline\hline
Run  \hspace{6mm} & $\epsilon$  \hspace{6mm} & $\theta_0$  \hspace{3mm} & 
$\eta$  \hspace{3mm} & $\beta$  \hspace{3mm} & $k_0$ & $m_0$ & 
${m_c}_\parallel$  & $m^-_\parallel$  & $\gamma$ \\
\hline
\tt{A0-Par} & 0.000 & $0^\circ$ & 0.2 & 0.1 & $2\pi$ & $10$ & $15$ & $5$ & $0.63$ \\
\tt{A1-Par} & 0.002 & $0^\circ$ & 0.2 & 0.1 & $2\pi$ & $10$ & $15$ & $5$ & $0.62$ \\
\tt{A2-Par} & 0.020 & $0^\circ$ & 0.2 & 0.1 & $2\pi$ & $10$ & $15$ & $5$ & $0.51$ \\
\tt{A3-Par} & 0.200 & $0^\circ$ & 0.2 & 0.1 & $2\pi$ & $10$ & $15$ & $5$ & $-$ \\
\tt{A0-Obl} & 0.000 & $30^\circ$ & 0.2 & 0.1 & $2\pi$ & $10$ & $16$ & $6$ & $0.39$ \\
\tt{A1-Obl} & 0.002 & $30^\circ$ & 0.2 & 0.1 & $2\pi$ & $10$ & $16$ & $6$ & $0.38$ \\
\tt{A2-Obl} & 0.020 & $30^\circ$ & 0.2 & 0.1 & $2\pi$ & $10$ & $16$ & $6$ & $0.27$ \\
\tt{A3-Obl} & 0.200 & $30^\circ$ & 0.2 & 0.1 & $2\pi$ & $10$ & $16$ & $6$ & $-$ \\
\tt{Nonmon} & 0.020 & $30^\circ$ & 0.2 & 0.1 & $2\pi$ & $10$ & $15$ & $5$ & $0.22$ \\
\hline
\end{tabular}
\label{table:runs}
\end{table}

Simulations are performed by employing the {\tt ECHO} code for classical, ideal MHD
where the EB has been implemented as described in Sect.~\ref{sect:model}.
We use $512\times 512$ grid points and high-order finite-difference methods, namely
a fifth order MPE5 upwind spatial reconstruction \citep{Del-Zanna:2007,Landi:2008}
and third order Runge-Kutta time integration, with a Courant number of 0.5.
The solenoidal constraint for the magnetic field is enforced through the Upwind Constrained
Transport (UCT) method \citep{Londrillo:2004}. All simulations are initialized by adding
a white noise of $10^{-4}$ density random fluctuations in order to trigger the instability.

Run A parameters, namely $\eta=0.2$ and $\beta=0.1$ \citep{Del-Zanna:2001, Del-Zanna:2001a},
will be chosen as default values, whereas different values of the expansion rate 
$\epsilon$ will be selected by choosing the pump wave period $\tau_A$. 
Both the parallel (circular polarization)
and oblique (arc polarization) cases will be studied. 
A last run will be devoted to a non-monochromatic (oblique) case. 
The list of parameters for each run is listed in Table~\ref{table:runs}.

Let us mention that runs with different sets of parameters 
have also been performed, in particular Run B ($\eta=0.5,\,\beta=0.5$) and 
Run C ($\eta=1,\,\beta=1.2$) in the cited papers. These values
were selected in order to obtain similar growth rates for the decay
instability and to mimic different conditions of the solar wind plasma,
namely those for an increasing heliocentric distance (here included
in the EB approach). 
Results are qualitatively similar, therefore we prefer not to show all of them here.
We would just like to remind that Run C parameters are those more
suitable to explain the observed saturation of the cross helicity to $\sigma\simeq 0.3$
at $\sim 2.5$~AU \citep{Bavassano:2000}.
As previously showed \citep{Del-Zanna:2012a}, even 1D runs in parallel propagation,
with an expansion rate $\epsilon=0.05$, seems adequate to reproduce the data.

\subsection{Parallel propagation of circularly polarized Alfv\'en waves}

Before presenting the effects of the solar wind expansion, we first
show an introductory run with $\epsilon=0$, to explain the
basics of parametric decay and define some important characteristic quantities.
Thus we start by studying the evolution of a parallel propagating pump Alfv\'en wave 
with $\theta_0=0$ and circular polarization (run {\tt A0-Par}).


\begin{figure}
\centerline{
 \includegraphics[height=5.0cm]{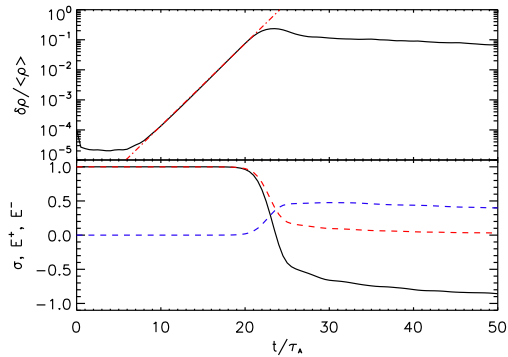}
  }
\caption{
Results from a simulation of parallel propagating Alfv\'en waves
and no expansion ($\epsilon=0$, run {\tt A0-Par}). 
The instability grows with a measured rate $\gamma=0.63$ 
(the dot-dashed fit line).
In the top panel we show the \emph{rms} density fluctuations as a function of time,
whereas in the bottom panel we plot the corresponding cross helicity and the Els\"asser
energies (dashed lines).
}
\label{fig:test}
\end{figure}


In Fig.~\ref{fig:test} we can follow the main indicators for the decay instability.
In the above panel we show the root mean square (\emph{rms}) density fluctuations, 
in logarithmic scale, as function of the time $t$. The linear phase of the decay instability
is clearly evident, and an estimation of the exponential growth  of density
fluctuations yields $\gamma=0.63$, close to the expected value 
of $\gamma = 0.11\times 2\pi= 0.69$ for the \emph{maximum} growth rate
predicted by Eq.~\ref{eq:dispersion}.
This $10\%$ error is partly due to the fact that in a periodic domain it is almost 
impossible to retrieve the most unstable mode, since only integer numbers of 
wavelengths are allowed, thus only selected wave numbers can be retrieved. 
Moreover, \emph{rms} quantities refer to a mixture of unstable modes, in our
case $m_c=15$ and $m_c=16$ (a spectral analysis would be required to 
single out the different modes), and even in parallel propagation 2D effects
may play a role (excitation of transverse modes, see below).
A fit to a 1D run, preserving the same resolution of 512 grid points, leads
to a slightly larger rate $\gamma=0.64$, to be compared with the theoretical expectations 
from Eq.~\ref{eq:dispersion} of $\gamma (m_c=15)=0.65$ and $\gamma (m_c=16)=0.66$ 
(we recall that the peak would be for $m_c\simeq 15.5$), corresponding to errors of $1.5-3\%$,
respectively. 

Additional diagnostics comes from the time dependency of Els\"asser energies.
In the bottom panel of Fig.~\ref{fig:test} we plot
\be
E^\pm = < \tfrac{1}{2} |\mathbf{z}^\pm |^2 >, \quad 
\mathbf{z}^\pm = \delta\mathbf{\varv} \mp \delta\mathbf{B}/\!\sqrt{\rho},
\ee
normalized against the initial $E^+$ value, together with
the cross helicity
\be
\sigma = \frac{E^+-E^-}{E^++E^-},
\ee
whose value is $\sigma=+1$ until the pump wave dominates and starts to
decrease when the instability peaks, around $t=20$ in this case.
When the compressive mode at $k_c>k_0$ has fully developed, shock
heating is known to take place, the instability saturates and density
fluctuations cease to increase. Correspondingly, the $E^+$ energy
decreases and the Alfv\'enic backscattered mode gains energy.
This process is strongly dependent on the value of the plasma $\beta$. 
In such a low-$\beta$ case we find that $\sigma$ even approaches $-1$, 
meaning that $E^+\to 0$ and an almost pure new pump wave $\mathbf{z}^-$ (with $k^-<k_0$) 
is present. As previously noticed, this situation leads to multiple decays 
at longer times \citep{Del-Zanna:2001}.

We start our analysis of the effects of the expansion by investigating the case
of parallel propagation ($\mathbf{k}_0 \parallel \mathbf{B}_0 \Rightarrow \theta_0 =0$) 
and circular polarization of the pump Alfv\'en wave. 
As for the test case without expansion, we employ
the usual parameters ($\eta=0.2$, $\beta=0.1$) and the numerical settings 
previously described. As far as expansion is concerned, we run here
three simulations with $\epsilon=0.002$ ({\tt A1-Par}), $\epsilon=0.02$ ({\tt A2-Par}), 
and $\epsilon=0.2$ ({\tt A3-Par}).
If we choose a constant wind speed $U_0=750$~km~s$^{-1}$ and an initial
position $R_0=0.25$~AU, then the three values span the whole Alfv\'enic range
for frequencies from $\nu_A=10^{-2}$~Hz ($\epsilon=0.002$) to $\nu_A=10^{-4}$~Hz
($\epsilon=0.2$), according to Eq.~\ref{eq:epsilon}.

The chosen values of $\epsilon$ must be compared with the expected instability
growth rate $\gamma=0.63$ for the parameters already employed
for {\tt A0-Par} in the absence of expansion. Since $\epsilon = U_0/R_0 = \dot{a}/a(0)$,
when this \emph{Hubble-like} term satisfy $\epsilon\ll\gamma$ only minor effects
are expected on the exponential growth of the instability, but when $\epsilon\,\ltsim\,\gamma$
the instability is strongly modified or even suppressed \citep{Del-Zanna:2012a}, and
if still present the growth in time becomes algebraic rather than exponential \citep{Tenerani:2013}.
This is not surprising: in the first case the pump wave can still be seen
as a quasi-stationary solution of the unperturbed problem, while for $\epsilon\,\ltsim\,\gamma$
it becomes a time-dependent solution, and the growth can not be any longer exponential.
The first two cases corresponding to $\nu_A=10^{-2}$~Hz and $\nu_A=10^{-3}$~Hz
belong to the first category, while pump waves at the lowest frequency limit
of the observed Alfv\'enic range, namely $\nu_A=10^{-4}$~Hz ($\epsilon=0.2$),
are expected to have non-exponentially growing daughter modes.


\begin{figure}
\centerline{
  \includegraphics[height=5.0cm]{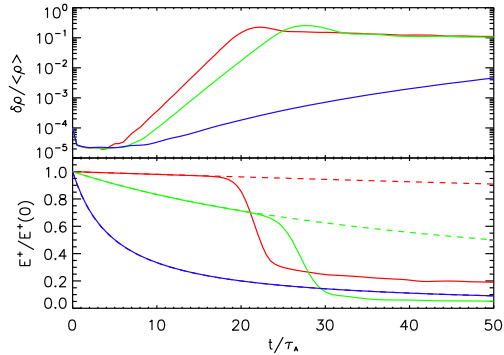}
  }
\caption{
Comparison of three simulations with mother wave in parallel propagation. 
Curves from top to bottom: $\epsilon=0.002$ ({\tt A1-Par}, red lines in the color version), 
$\epsilon=0.02$ ({\tt A2-Par}, green), and $\epsilon=0.2$ ({\tt A3-Par}, blue). 
In the top panel density fluctuations are shown as a function of time, 
whereas in the bottom panel we plot the normalized energy associated to $\mathbf{z}^+$,
compared with the expected $E^+\sim [a(t)]^{-1}$ evolution in the absence of instability 
(dashed curves).
}
\label{fig:comparison}
\end{figure}



\begin{figure}
\centerline{
 \includegraphics[height=5.5cm]{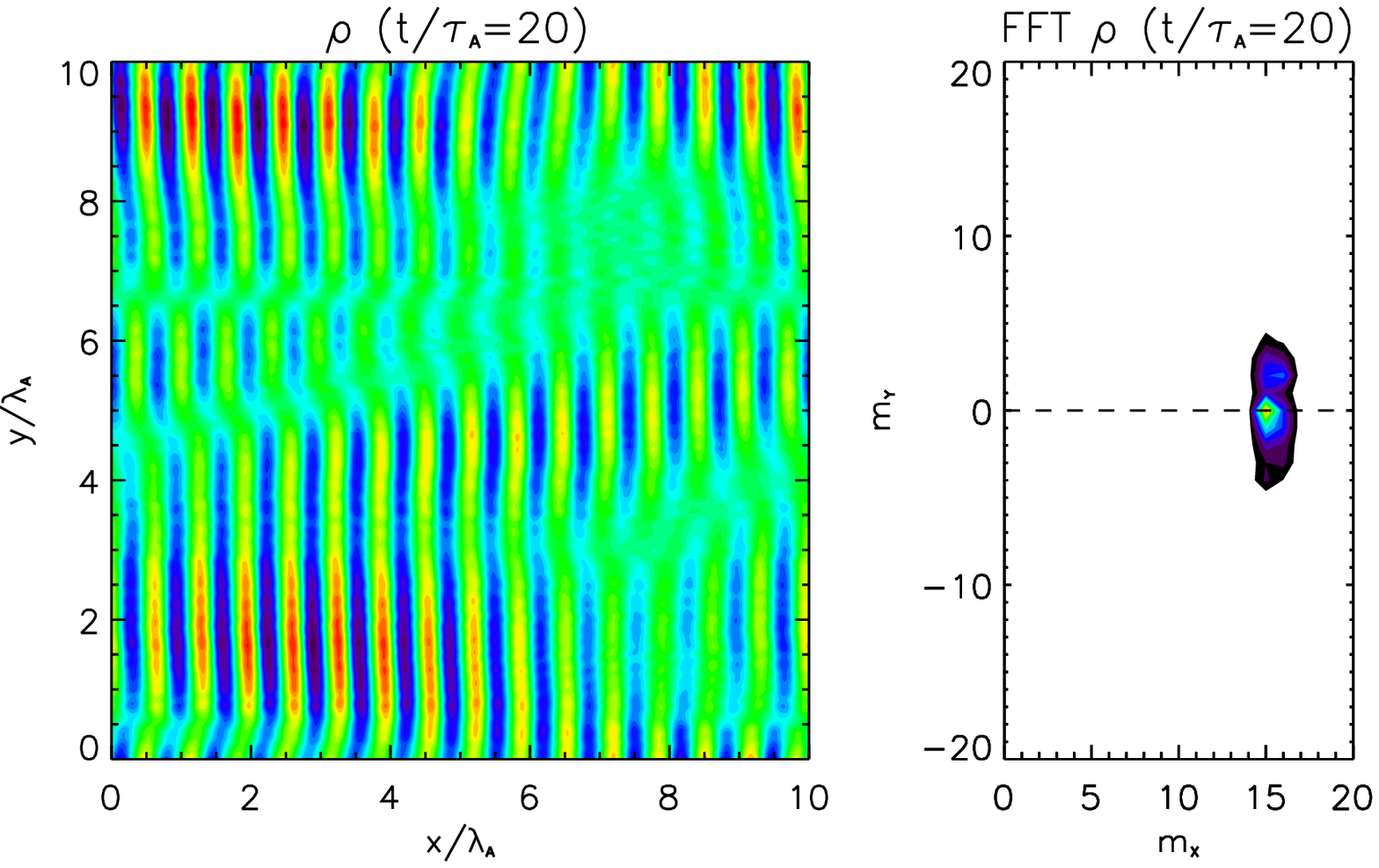}
 \includegraphics[height=5.5cm]{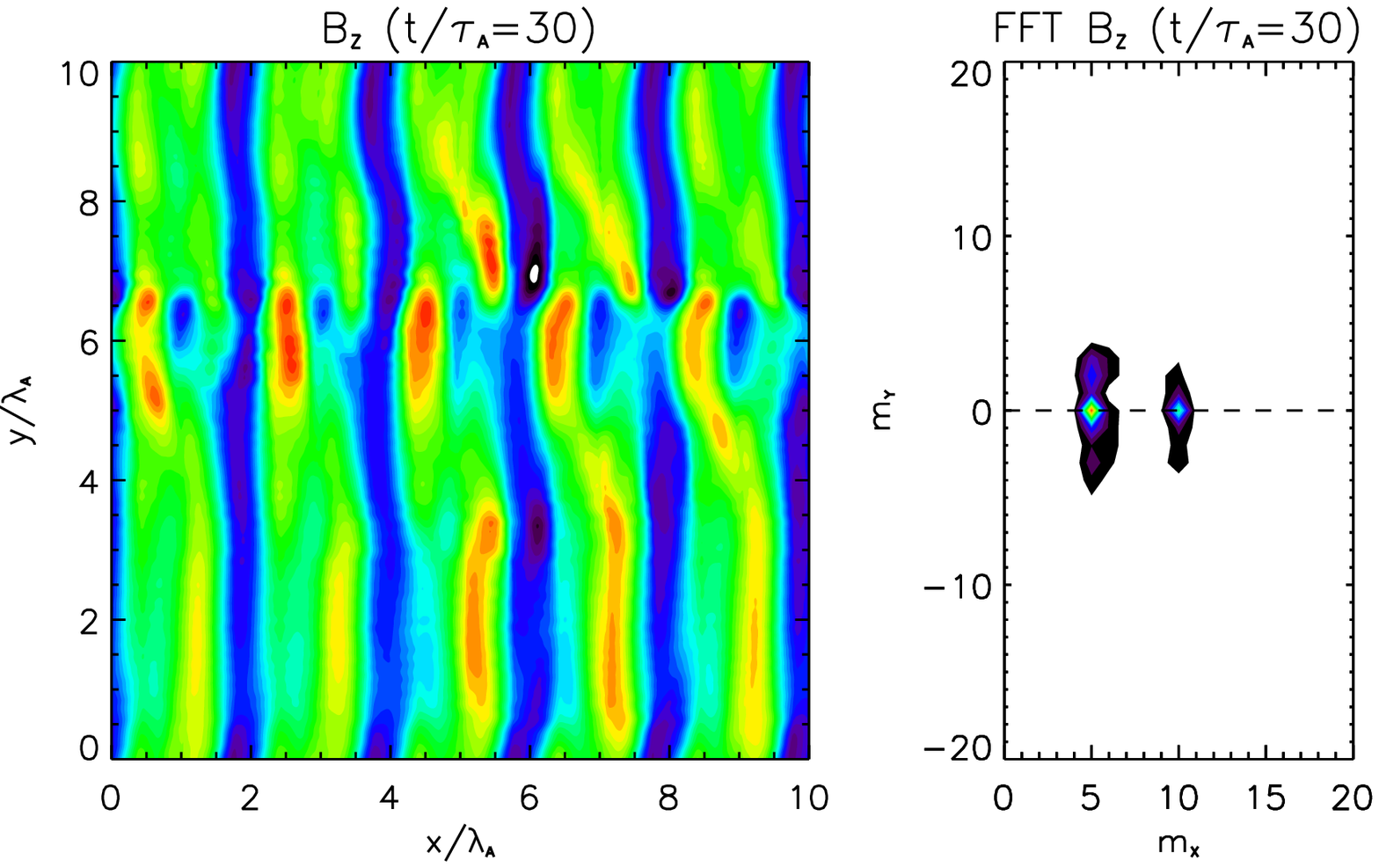}
   }
\caption{
2D plots and spectra of density at $t=20 \, \tau_A$ (left panels) and 
transverse magnetic component $B_z$ at $t=30 \, \tau_A$ (right panels) 
for the parallel propagation case with $\epsilon=0.02$ ({\tt A2-Par}).
Quantities are displayed in the comoving frame (without stretching along $y$).
The compressive mode with $m_x=15$ and the backscattered
Alfv\'enic mode with $m_x=5$ are clearly visible. The horizontal
dashed line in the FFT plots indicates the radial direction of 
$\mathbf{B_0}\parallel \mathbf{k}_0$.
}
\label{fig:parallel}
\end{figure}


In Fig.~\ref{fig:comparison} we show the comparison of the quantities characterizing
the evolution of the decay instability for the three runs. In the top panel we plot the
density \emph{rms} fluctuations (normalized to the local average density at every time),
it is evident that for $\epsilon=0.002$ the effects of the wind expansion are basically
negligible (see Fig.~\ref{fig:test} for further comparison), for $\epsilon=0.02$ the
onset of the instability is slightly delayed and the growth rate is somehow smaller,
while for $\epsilon=0.2$ the effects of the expansion are so strong that the instability
is almost suppressed. As predicted, in the latter case of the longest period pump wave,
the slow growth of density fluctuations is far from being exponential. Again, this is due
to the action of the $\dot{a}/a$ terms in the equations (see the source terms in the original
formulation), which are strongly and continuously modifying the background
plasma parameters before but even during the onset of the instability. 
In the bottom panel we show the energy $E^+$ of the forward propagating
Alfv\'en mode $\mathbf{z}^+$, normalized to its initial value.
The expected decay, purely due to the expansion of the medium in which the
pump wave propagates, is indicated with dashed lines, for each run.
Notice that for the small $\epsilon=0.002$ value the evolution is almost identical
to the non-expanding case. 
In the intermediate case the expansion affects the evolution, but $E^+$ drops
considerably due to the instability, in spite of the gradual reduction due to
the $\dot{a}/a$ terms. For the extreme $\epsilon=0.2$ ({\tt A3-Par}) case the expansion
effects are so strong that no contribution to its decay due to the instability can
be seen in the reported range of times.
Moreover, we should recall here that quantities have been plotted in terms
of time normalized to the pump wave period, so that for increasing $\epsilon$
the evolution is actually slower, in absolute terms, also due to the increment
of the normalization period $\tau_A$.

Let us now investigate the 2D properties of the decay instability in the presence
of wind expansion and consider the intermediate $\epsilon=0.02$ case ({\tt A2-Par}).
In Fig.~\ref{fig:parallel} we plot in the equatorial $x-y$ plane
the density $\rho$ at $t=20\tau_A$, just before the saturation of the
instability, and magnetic component $B_z$ at $t=30\tau_A$, just after saturation, 
representative of compressive and Alfv\'enic modes, respectively.
Notice that the periodic box has selected
a main $m_x=15$ magneto-acoustic mode along the propagation direction of 
$\mathbf{k}_0\parallel\mathbf{B}_0$, though a lower wave number modulation 
along $y$ is also present. At this time the sound-like waves are about
to steepen into a train of shocks, though the mother pump wave is still dominant.
As far as the Alfv\'enic component is concerned, we then need to wait a later
time after saturation (occurring around $t=25\tau_A$, see Fig.~\ref{fig:comparison}) 
to see a dominant daughter wave. We then plot the $B_z$ component
at $t=30\tau_A$, when the $\mathbf{z}^+$ pump wave has basically died and
only the backscattered $\mathbf{z}^-$ wave is visible, with $m_x=5$ and an additional 
low wave number modulation in the $y$ direction.

The same information can be more easily drawn by investigating the reported 
2D spectra, obtained through \emph{Fast Fourier Transforms} (FFT) of the selected
physical quantities. The daughter waves are clearly peaked at mode numbers
$\mathbf{m}=(15,0)$ for the compressive component and at $\mathbf{m}=(5,0)$ for the 
Alfv\'enic mode, though \emph{stripes} of excited modes in the $y$ direction are also
present at low $m_y$ numbers. 
We should remind here that parametric decay of pump waves in parallel
propagation is essentially a 1D process,
so we did not expect to see a large scatter in the 2D spectra.
This fact is clearly reflected by the presence
of daughter waves that peak in the FFT plots mainly along the dashed line,
indicating the radial direction of both $\mathbf{B_0}\parallel \mathbf{k}_0$.

\subsection{Oblique propagation of arc-polarized Alfv\'en waves}

The situation is quite different in the oblique case.
We repeat the same three runs ($\epsilon=0.002$, $\epsilon=0.02$, $\epsilon=0.2$)
with Run A parameters, this time with an initial angle $\theta_0=30^\circ$ between
the pump wave vector $\mathbf{k}_0$ in the radial direction and the background
field $\mathbf{B}_0$. The Alfv\'en wave polarization is now of arc-type, as described
in Sect.~\ref{sect:init}, necessary to preserve an overall $|\mathbf{B}|=\mathrm{const}$
during the evolution and thus to avoid ponderomotive forces that may spoil the development
of the decay instability.

\begin{figure}
\centerline{
 \includegraphics[height=5.0cm]{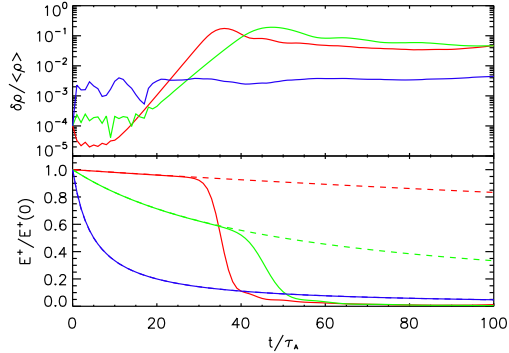}
  }
\caption{
Comparison of three simulations with mother wave in oblique propagation. Curves from
top to bottom: $\epsilon=0.002$ ({\tt A1-Obl}, red lines in the color version), 
$\epsilon=0.02$ ({\tt A2-Obl}, green), and $\epsilon=0.2$ ({\tt A3-Obl}, blue). 
Notations are as in Fig.~\ref{fig:comparison}.
}
\label{fig:obl_comparison}
\end{figure}

In Fig.~\ref{fig:obl_comparison} we plot the comparison of the quantities
characterizing the evolution of the instability for the three runs with different
expansion factors $\epsilon$, in analogy to the parallel case already
shown in Fig.~\ref{fig:comparison}. Notice that the overall evolution
is now always slower, as shown in \citet{Del-Zanna:2001a}, with
correspondingly smaller growth rates. We recall that in the oblique
case and arc polarization we expect a dependence $\gamma\sim\cos\theta_0$,
though for $\theta_0=0$ we cannot recover the growth rate for parallel
propagation.
However, the qualitative behavior of the \emph{rms} quantities is very
similar to the case of parallel propagation. The main differences are
a somehow stronger decrease of the $E^+$ energy after saturation,
simply due to the later occurrence of the instability, so that the pump wave
amplitude is already smaller due to the wind expansion, and a stronger
growth of the initial density noise, that completely prevents any sign
of instability onset in the extreme expansion case $\epsilon=0.2$.


\begin{figure}
\centerline{
 \includegraphics[height=5.5cm]{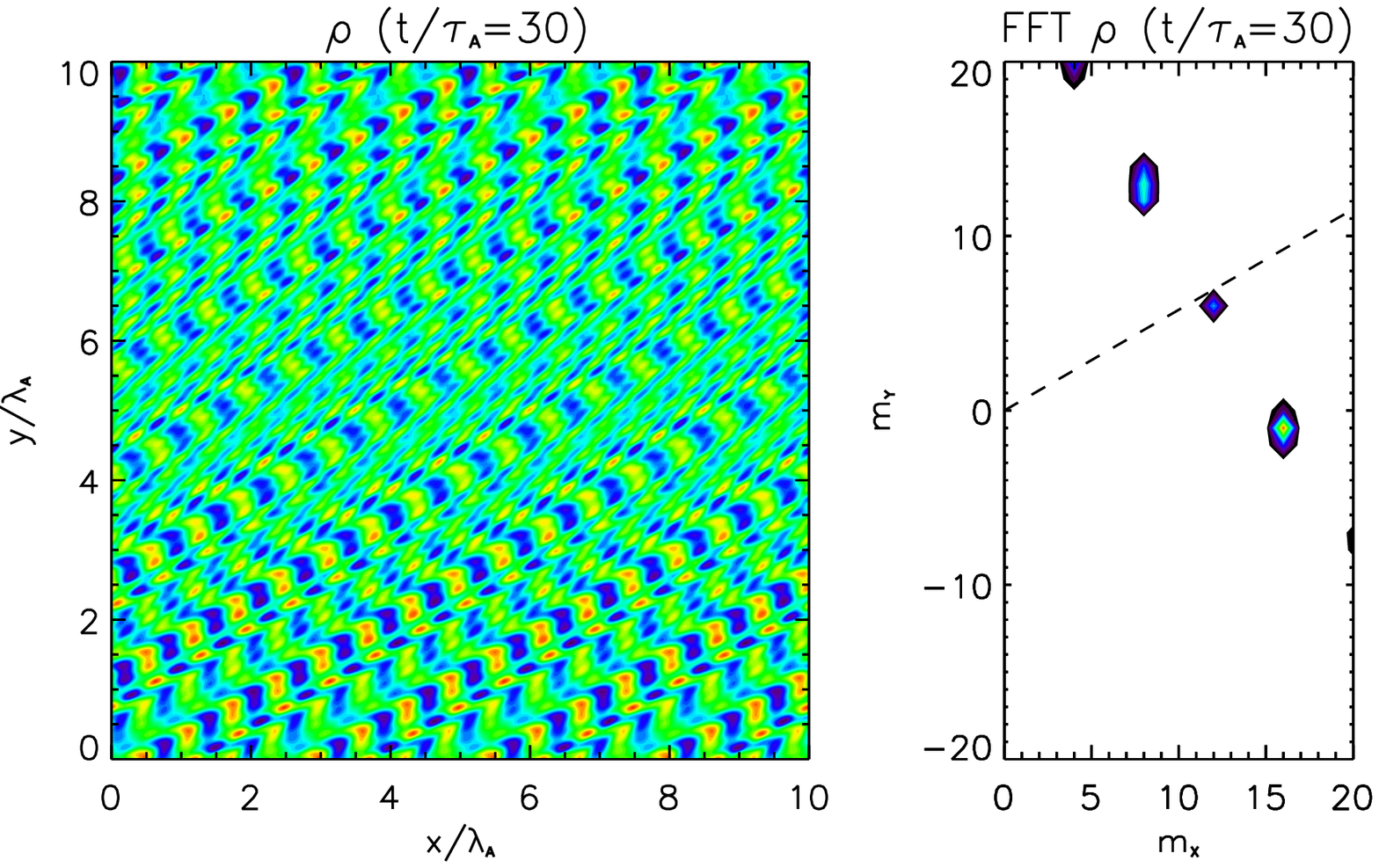}
  \includegraphics[height=5.5cm]{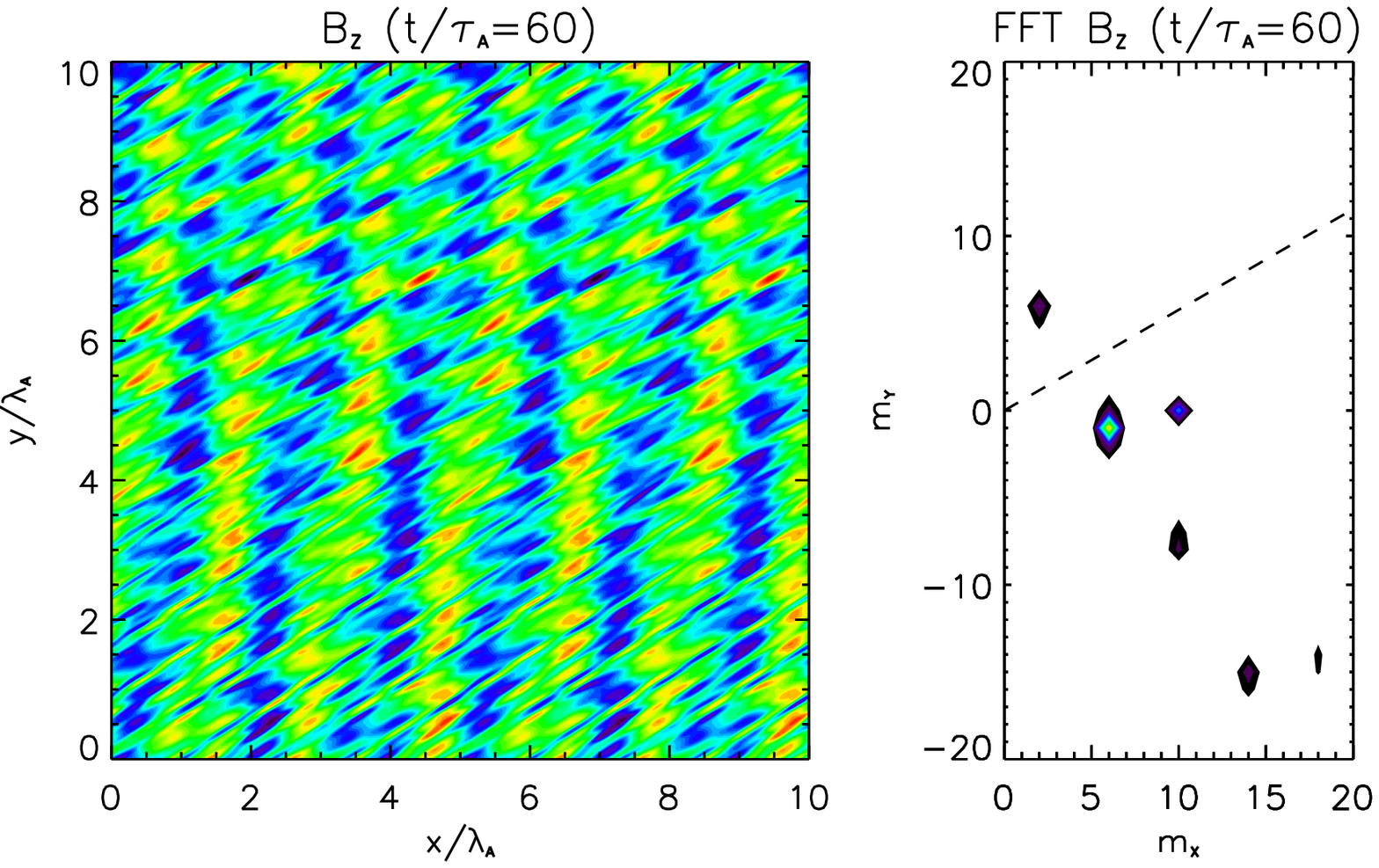}
   }
\caption{
2D plots and spectra of density at $t=30 \, \tau_A$ (left panels) and 
transverse magnetic component $B_z$ at $t=60 \, \tau_A$ (right panels) 
for the oblique propagation case {\tt A2-Obl}, in the comoving frame.
The dashed line in the FFT plots indicates the initial inclination of
$\theta_0=30^\circ$ for $\mathbf{B}_0$.
}
\label{fig:oblique}
\end{figure}


The 2D behavior of fluctuations is more interesting in the oblique case.
When the decay occurs in 2D two resonance conditions must be satisfied, namely
\be
{k_0}_\parallel={k_c}_\parallel - k^-_\parallel, \quad
{k_0}_\perp={k_c}_\perp - k^-_\perp .
\label{eq:k}
\ee
Since the mother wave propagates along $x$ but the
background magnetic field (the parallel direction) is initially inclined with 
$\theta=30^\circ$, ${k_0}_\perp\neq 0$ and a direct excitation
of transverse modes in the daughter compressive and reflected Alfv\'enic waves
is expected. This is clearly visible in Fig.~\ref{fig:oblique}, referring to the
intermediate case {\tt A2-Obl}: a compressive mode
with a dominant ${m_c}_x = 16$ has developed, with fronts no longer aligned
along $x$ (the direction of $\mathbf{k}_0$), but roughly along the direction
of $\mathbf{B}_0$. A strong modulation is seen in the perpendicular direction too,
even at high wave numbers, which was absent in the case of parallel propagation.
Similarly, the 2D contours of $B_z$ at $t=60\tau_A$ (after saturation, when $E^->E^+$) 
show the presence of a $m^-_x=6$ dominant mode, with a mixture of positive
and negative $m^-_y$ excited modes and again a strong modulation
in the transverse direction too.

The situation is clearer by inspecting the corresponding FFT plots.
Since the decay channels are larger in a 2D geometry, allowing
more nonlinear interactions, a broadband spectrum of daughter
waves is expected. In Fig.~\ref{fig:oblique} we notice that while
only a single value of ${k_c}_\parallel$ (and ${k^-}_\parallel$) is
excited, a range of oblique modes with different ${k_c}_\perp$ 
(and $k^-_\perp$), all sharing the same $k_\parallel$, can be clearly
seen in the form of oblique \emph{dotted stripes} in the FFT plots.
Here we iterate that the presence of these additional excited waves
by mode coupling is due to the fact that ${k_0}_\perp\neq 0$ right from the start,
then also (at least one of) the daughter waves must have a perpendicular
wave vector component.

This 2D behaviour is not peculiar of MHD only, it was actually first observed and 
discussed in the hybrid regime, for \emph{linearly} polarized oblique waves,
where the main difference is that the instability saturation occurs earlier and 
the mechanism is different (proton trapping).
Formation of a proton beam is always observed along $\mathbf{B}_0$,
even for oblique propagation, and the electron to proton temperature ratio also plays 
an important role \citep{Matteini:2010a,Del-Zanna:2012}.


\begin{figure}
\centerline{
     \includegraphics[height=5.0cm]{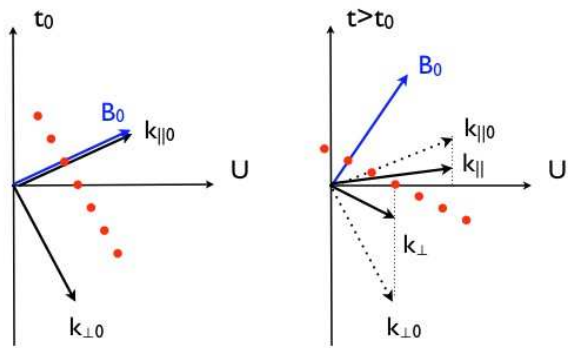}
  }
\caption{
Modifications in physical space due to expansion of the background 
field $\mathbf{B}_0$ and of the wave vector $\mathbf{k}_0$ 
(the decomposition refers to the field position at time $t_0$).
}
\label{fig:obl_sketch}
\end{figure}


As far as the effects of expansion are concerned, 
the situation is that outlined in the sketch of Fig.~\ref{fig:obl_sketch},
here in \emph{physical} space, where the lateral stretching in $y\sim a$
and the shrinking in $k_y\sim a^{-1}$ are both taken into account.
The inclination of $\mathbf{B}_0$ increases in time
(the development of the Parker spiral in the $x-y$ equatorial plane), whereas
wave vectors tend towards the (radial) $x$ axis, as $k_y$ decreases.
The net result is that any  $\mathbf{k}$ which is originally perpendicular
to $\mathbf{B}_0$ will always be perpendicular also to the rotated field.
On the other hand, the component $\mathbf{k}_\parallel$, initially
parallel to  $\mathbf{B}_0$ by definition, diverges from that direction
as time increases leading to $\mathbf{k}_\parallel \nparallel \mathbf{B}_0$.
It is now easy to understand why the stripe of daughter waves
always remains perpendicular to $\mathbf{B}_0$: the creation
of these modes occurs on a timescale which is much shorter than
the expansion, so that the subsequent evolution is only kinematical
following the EB stretching, as described just above.
If we finally return to Fig.~\ref{fig:oblique}, notice that the stripe
of transverse modes does not rotate, since its direction is frozen in this
\emph{comoving} Fourier space, and this is always perpendicular
to the initial $\mathbf{B}_0$ direction.
Similar effects are observed in 2D and 3D applications of EB 
to MHD turbulence evolution studies \citep{Grappin:1993,Grappin:1996,Dong:2014}.

Notice that here we have shown the results for a moderate value of the expansion
factor ($\epsilon=0.02$), but we would like to stress that the direct excitation of
transverse modes is simply a consequence of the oblique propagation of the
pump wave in a 2D domain, allowing for extra channels available for decay.
Very similar results are obtained with $\epsilon=0.002$ (or even 
$\epsilon=0$), whereas for $\epsilon=0.2$ the expansion is too
strong and prevents the onset of the decay, as shown previously.

\subsection{Non-monochromatic oblique case}
\label{sect:nonmon}

What shown so far refers to the simple case of a monochromatic
pump wave. While this is important to investigate the decay instability
itself, free from other spurious effects, the model problem is
certainly far from being realistic.
A superposition of pump Alfv\'enic modes could be studied, of course,
but the problem is that the strength of the magnetic fluctuations 
do not preserve the condition $|\mathbf{B}|=\mathrm{const}$,
so the initial ponderomotive force drives compressive modes
that mask those arising from the decay instability that we want to study. 
A trick to reconcile these two requirements is due to \citet{Malara:1996}. 
While this and subsequent  works dealt with 1D parallel propagation
and circular polarization, here we apply the method for 2D oblique
polarization of arc-type pump waves for the first time.
The idea is to preserve the initial settings exactly as described
in Sect.~\ref{sect:init} for any kind of polarization, with no superposition
of individual waves, but to introduce multiple modes by modifying
the phase $\varphi$. We choose the form
\be
\varphi = k_0x + A \sum_{m=1}^{100} \frac{1}{m} \cos(k_m x+\varphi_m),
\label{eq:spectrum}
\ee
where $\varphi_k$ is a random phase for each value of $k_m=2\pi (m/m_0)$ 
(we recall that $k_0=2\pi$) and we set $A=1$. 
With this setting the novel $\mathcal{F}(\varphi)$ still provides 
an arc-polarized Alfv\'en wave where the mode number $m_0=10$ is again
the strongest, with tails at both lower and higher values of $m$, each mode
having a power decreasing as $m^{-2}$.


\begin{figure}
\centerline{
    \includegraphics[height=5.0cm]{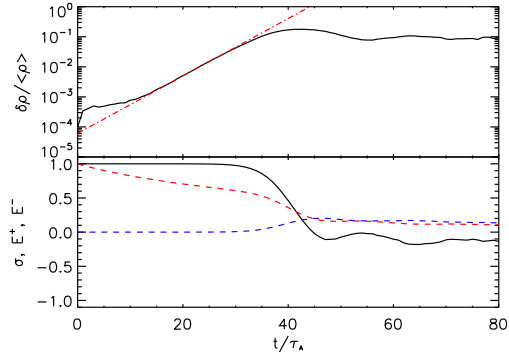}
  }
\caption{
The characteristic \emph{rms} quantities for parametric decay
for the non-monochromatic case with $\epsilon=0.02$ (run {\tt nonmon}).
The measured instability growth rate is now $\gamma=0.22$.
}
\label{fig:nonmon_rms}
\end{figure}


We repeat the simulation for the most interesting case
of $\nu_A = 10^{-3}$~Hz, corresponding to an expansion factor $\epsilon=0.02$
(we indicate this run as {\tt nonmon}, with the same settings as {\tt A2-Obl} 
other than the phase $\varphi$).
In Fig.~\ref{fig:nonmon_rms} we show the usual \emph{rms} quantities (normalized
density fluctuations, normalized Els\"asser energies, and cross helicity) and
we can clearly see that the decay instability takes place, though the growth rate
is reduced down to $\gamma=0.22$. The final state is not characterized
by a dominance of $\mathbf{z}^-$ modes as in the previous cases with the same
parameters, but we see that $E^-\sim E^+$ with oscillations of the cross helicity 
about $\sigma\sim  0$, though with negative values.


\begin{figure}
\centerline{
  \includegraphics[height=5.5cm]{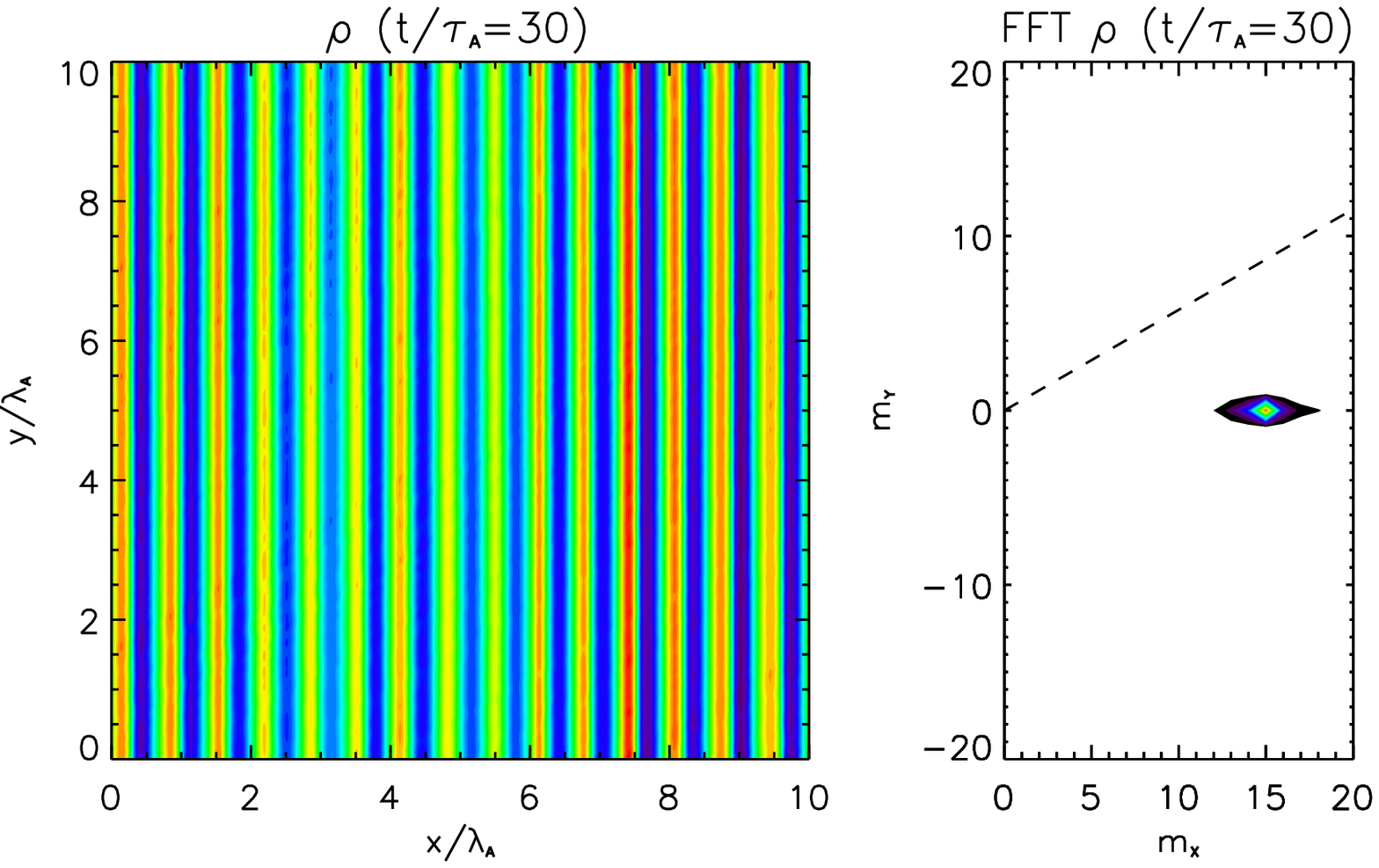}
 \includegraphics[height=5.5cm]{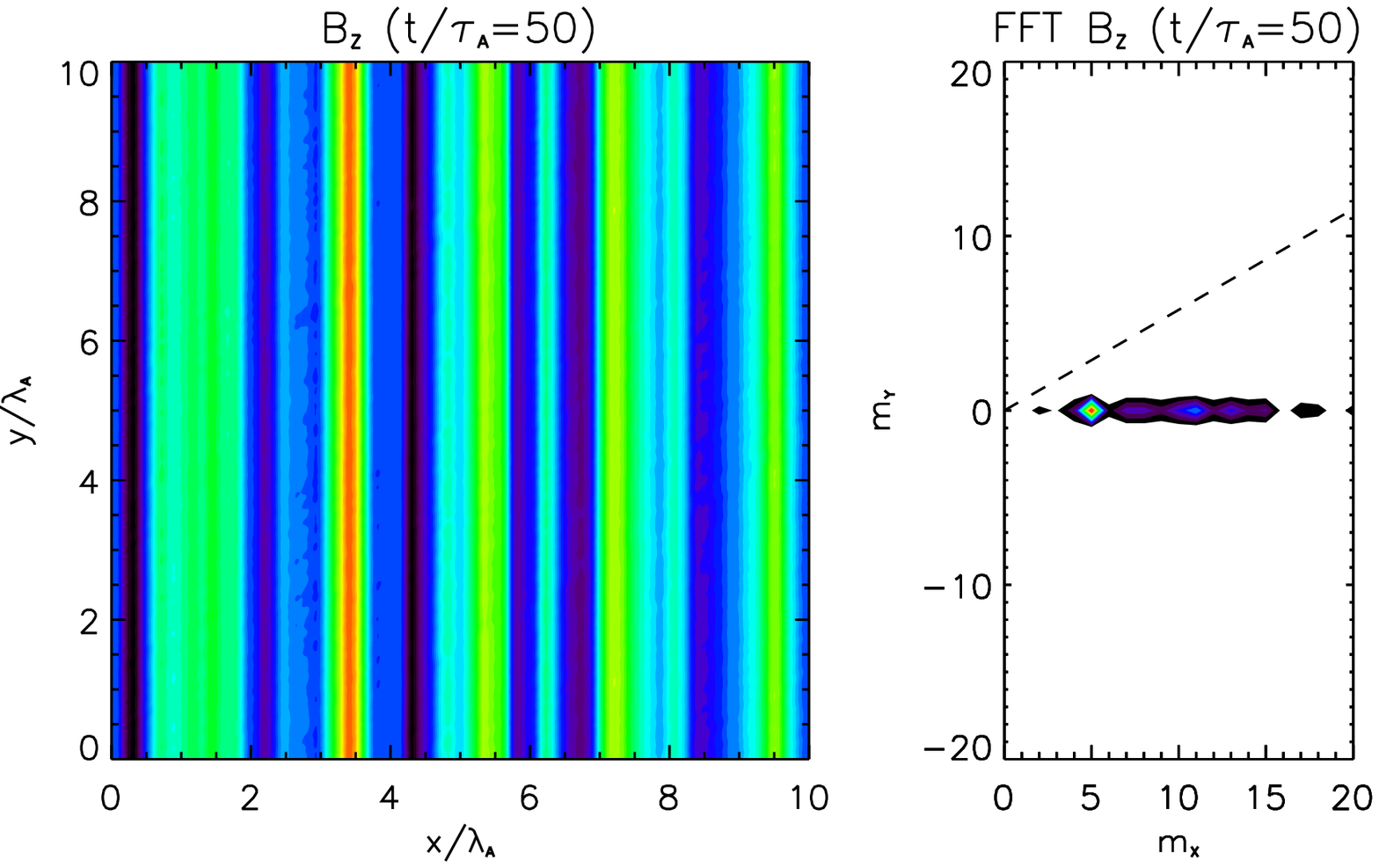}
   }
\caption{
2D plots and spectra of density at $t=30 \, \tau_A$ (left panels) and 
transverse magnetic component $B_z$ at $t=50 \, \tau_A$ (right panels) 
for the non-monochromatic case (run {\tt nonmon}).
}
\label{fig:nonmon_oblique}
\end{figure}


What is somehow unexpected is the 2D behavior of the instability. 
In Fig.~\ref{fig:nonmon_oblique}
we report, as usual, the 2D plots and spectra of the density fluctuations just
before the peak of the instability (left panels) and of the $B_z$ magnetic component
right after saturation (right panels). Clearly, no modulation at all is visible
in the $y$ direction, in spite of the oblique inclination of the background field
of $30^\circ$ (constant in time in Fourier space, as shown by the dashed lines).
On the other hand, the daughter waves are well characterized by $\mathbf{m}=(15,0)$
for $\delta\rho$ and by $\mathbf{m}=(5,0)$ for the backscattered Alfv\'en wave.
A larger spread in $m_x$ is observed in the $B_z$ spectrum, due to the
presence of the pump wave, which has not decayed to $E^+\ll E^-$ values,
characterized by a dominant $\mathbf{m}=(10,0)$ mode with additional
components as from Eq.~\ref{eq:spectrum}.
The reason for this behavior is that the initial broad-band spectrum of the 
mother wave imposes much stringent phase relations to be satisfied by the
daughter waves as compared to the monochromatic case \citep{Matteini:2012a}.
Resonances are then more difficult to excite perpendicularly
to the initial $\mathbf{k}_0$, parallel to the $x$ axis, and the cascade only
occurs along that direction, precisely as in a purely 1D case.

\section{Conclusions}
\label{sect:conclusions}

We presented, for the first time, a detailed analysis of the solar wind 
expansion effects  in two-dimensional MHD simulations of the parametric 
decay instability of Alfv\'en waves, for both parallel and oblique propagation.
We took advantage of the expanding box (EB) model, here reformulated
geometrically by introducing a time-dependent metric tensor $\mathrm{diag}\{1,a^2,a^2\}$,
where $a(t)$ is the \emph{Hubble-like} transverse expansion factor.
We considered Alfv\'en waves of low frequency ($\nu\sim 10^{-4}-10^{-2}$~Hz)
as observed in the solar wind, and studied their nonlinear evolution
far enough from the Sun, where the solar wind speed $U_0$ can be considered 
constant. 

The background plasma parameters (namely the fluctuations normalized amplitude 
$\eta\sim a^{1/2}$ and the plasma $\beta\sim a^{2/3}$) are modified before
and during the onset of the instability, typically reducing the expected growth rate,
though for extreme values of $\epsilon$ the parametric instability may be
completely suppressed.
For our choice of parameters, appropriate for the inner heliospheric region 
($R_0=0.25$~AU, $U_0=750$~km~s$^{-1}$, $\eta=0.2$, $\beta=0.1$),
the instability increases very slowly (algebraically) for the mother Alfv\'en waves with the 
longest period $\tau_A=10^4$~s, corresponding to an expansion rate $\epsilon=0.2$,
which is comparable to the growth rate measured in the absence of expansion
($\gamma\simeq 0.63$ for the parallel propagation case).
On the other hand, for $\tau_A=10^3$~s ($\epsilon=0.02$) there is competition
between the mother wave decay due to the expansion and that due to the
instability, leading to a $15\%$ reduction of $\gamma$ but leaving the evolution
of quantities in $t/\tau_A$ qualitatively unchanged, whereas for  $\tau_A=10^2$~s 
($\epsilon=0.002$) the instability onset and evolution are not substantially
affected by the expansion. 
In the oblique case, with arc-polarized mother waves, the growth rates are always smaller 
than the corresponding parallel ones.

For a given expansion rate (the intermediate one $\epsilon=0.02$,
corresponding to a frequency $\nu_A=10^{-3}$~Hz), we
investigated the 2D behavior of the compressive and backscattered
Alfv\'enic daughter waves and the spectral properties of such fluctuations.
In the parallel case the excited modes lead to $m_c=15$ and $m^-=5$ wavelengths
in the numerical box along the $x$ direction of $\mathbf{k}_0$ (the pump wave has $m_0=10$), 
with a very little spread in Fourier modes for low $m_y$ transverse modes.
The situation changes in the corresponding oblique case.
Here we found $m_c=16$ and $m^-=6$ along $x$ for the main excited modes,
with $m_y=-1$, and other resonant daughter waves are also triggered.
Related to the 2D behavior of the parametric decay in the oblique case,
two novel results were found in the present work. 

First, the stripes of daughter
waves seen in Fourier space always form perpendicular to the local $\mathbf{B}_0$,
as first observed by \citet{Matteini:2010a} in hybrid simulations (without expansion).
When the solar wind expansion is taken into account, this field rotates
in time in the equatorial plane by increasing the $\theta$ angle with respect
to the radial direction, mimicking the presence of the Parker spiral.
However, this effect is compensated by the shrinking of the non-radial  
$\mathbf{k}$ components, thus  the stripe of daughter modes, which is created
and then remains frozen with the expansion, is always perpendicular to $\mathbf{B}_0$.

Second, if an initial non-monochromatic spectrum is set up for the mother wave
(by changing just the phase in order to preserve the condition 
$|\mathbf{B}_0+\delta\mathbf{B}|=\mathrm{const}$), the triggering of modes transverse
to the local $\mathbf{B}_0$ is no longer found, and all modes are excited along $x$,
with no sign of any modulation along $y$. This is probably explained by the more
stringent requirements for $\mathbf{k}$ and $\omega$ that the daughter waves 
should satisfy for all the excited modes and not just for those arising from
the dominant $m_0=10$ one. 
If this result will be confirmed by more general simulations (in full three dimensions
and allowing for a spectrum not restricted to the condition of a constant
$|\mathbf{B}_0+\delta\mathbf{B}|$), it would definitely mean that parametric
decay is unable to trigger a direct cascade of perpendicular modes, other
than just due to purely geometrical effects (at larger distances $\mathbf{k}_0$
and $\mathbf{B}_0$ tend to be progressively more perpendicular).

A final important aspect one should discuss is the possible relation between the
decay instability of low-frequency Alfv\'enic modes and the MHD turbulent cascade,
since both features are contemporarily  present in the heliospheric plasma.
The observations show that the magnetic fluctuations dominate
the low-frequency spectrum and are characterized by a $\nu^{-1}$ dependence,
probably reminiscent of the injection in the solar corona \citep{Verdini:2012a}.
At higher frequencies the spectrum is steeper, with
the typical Kolmogorov scaling $\nu^{-5/3}$ of fully developed turbulence.
The spectral break between the energy-containing scales and the inertial range, 
however, is seen to depend on the heliocentric distance, and in fast streams
this goes roughly as $\nu_\mathrm{b}\simeq 5\times 10^{-2}\,\mathrm{Hz}\, 
(R/0.3\,\mathrm{AU})^{-1.5}$ \citep[e.g.][]{Bruno:2013}.

We speculate that the solar wind expansion may act as a filter
for parametric decay and possibly explain the observed behavior
\citep[see also][]{Tenerani:2013}.
Waves with $\nu_\mathrm{A}\sim 10^{-2}-10^{-3}$~Hz, so with periods
of minutes, are basically insensitive to the solar wind expansion. 
Parametric decay is then free to operate by producing $\mathbf{z}^-$
modes out of the initial $\mathbf{z}^+$ ones, thus possibly triggering
a turbulent cascade, at least in the original radial direction of the pump 
wave vector (also parallel to the background field in 1D or oblique
to it in 2D). Mother waves with periods of hours,
on the other hand, feel the expansion effects, the $\epsilon$ parameter
is close to the expected parametric decay growth rate, and the
instability is strongly delayed or even suppressed.
Waves with frequencies of $\nu_\mathrm{A} \sim 10^{-4}$~Hz  
are thus not expected to evolve,
and this seems to be confirmed by the Ulysses observations
up to to 4.8~AU, reported also in the cited review, showing that at these 
frequencies we are still above the spectral break, where the
turbulent cascade has not been activated yet.
In any case, provided that the simple creation of $\mathbf{z}^-$ radially
incoming waves is really enough to trigger the cascade, then
the observed decrease of $\nu_\mathrm{b}$ with $R$ would be
naturally explained, as at larger distances (longer times), 
smaller and smaller frequency modes are expected to become 
parametrically unstable.

These aspects obviously would require a deeper understanding and a quantitative
comparison with real data, as well as the possible relation between parametric
decay of oblique waves and the transverse turbulent cascade should be studied
carefully in detail. However, we deem that these subjects are beyond the goal
of the present paper, and we leave more quantitative estimations and further
discussions as future work.

\acknowledgements 
The authors thank Roland Grappin for stimulating discussions and two anonymous referees
for their useful suggestions.
The research leading to these results has received funding from the European Commission's Seventh
Framework Program (FP7/2007-2013) under the grant agreement SHOCK (project number 284515).
The research described in this paper was also supported by the UK Science and 
Technology Facilities Council grant ST/K001051/1, and by a grant from the Interuniversity Attraction 
Poles Programme initiated by the Belgian Science Policy Office (IAP P7/08 CHARM).
This work was carried out in part at the Jet Propulsion Laboratory  under a contract with NASA. 
We also acknowledge support from the Italian Space Agency.

\bibliographystyle{jpp}


\end{document}